\documentclass[twocolumn,showpacs,amsmath,amssymb,aps,prl]{revtex4}
\usepackage[dvips]{graphicx}
\def\be{\begin{equation}}
\def\ee{\end{equation}}

\def\bea{\begin{eqnarray}}
\def\eea{\end{eqnarray}}

\def\e{\epsilon}

\def\s{\sigma}

\def\p{\varphi}

\def\Tr{{\rm Tr}}
\def\Ma{\check{a}}
\newcommand{\ev}[1]{\mbox{$\langle #1 \rangle$}}

\makeatletter
\def \underbracket {%
\@ifnextchar [{\@underbracket }{\@underbracket [\@bracketheight ]}%
}
\def \@underbracket [#1]{ %
\@ifnextchar [{\@under@bracket [#1]}{\@under@bracket [#1][0.4 em ]}%
}
\def \@under@bracket [#1][#2]#3{ %\ message { Underbracket : #1 ,#2 ,#3}
\mathop {\vtop {\m@th \ialign {##\crcr $\hfil\displaystyle {#3}\hfil $
\crcr \noalign {\kern 3\p@ \nointerlineskip }\upbracketfill {#1}{#2}
\crcr \noalign {\kern 3\p@ }}}}\limits }
\def \upbracketfill #1#2{$\m@th \setbox \z@ \hbox {$\braceld $}
\edef \@bracketheight {\the \ht\z@ }\bracketend {#1}{#2}
\leaders \vrule \@height #1 \@depth \z@ \hfill
\leaders \vrule \@height #1 \@depth \z@ \hfill \bracketend {#1}{#2}$}
\def \bracketend #1#2{\vrule height #2 width #1\relax }
\makeatother
\begin{document}
\pagestyle{empty}
%\preprint{MANUSCRIPT}

\title{Evolution of entanglement entropy following a quantum quench: \\
Analytic results for the XY chain in a transverse magnetic field}

\author{Maurizio Fagotti} \author{Pasquale Calabrese}
\affiliation{Dipartimento di Fisica dell'Universit\`a di Pisa and INFN, 
Pisa, Italy}

\date{\today}

\begin{abstract}

The non-equilibrium evolution of the block entanglement entropy is 
investigated in the XY chain in a transverse magnetic field after the 
Hamiltonian parameters are suddenly changed from and to arbitrary values. 
Using Toeplitz matrix representation and multidimensional phase methods, we
provide analytic results for large blocks and for all times, showing
explicitly the linear growth in time followed by saturation. 
The consequences of these analytic results are discussed and the
effects of a finite block length is taken into account numerically.

\end{abstract}


\maketitle

The non-equilibrium evolution of extended quantum systems is one of the most
challenging problems of contemporary research in theoretical physics. 
The subject is in a renaissance era after the experimental realization 
\cite{exp1} of cold
atomic systems that can evolve out of equilibrium in the absence of any
dissipation and with high degree of tunability of Hamiltonian parameters.
A strongly limiting factor for a better understanding of these phenomena is
the absence of effective numerical methods to simulate the dynamics of 
quantum systems. For methods like time dependent density matrix
renormalization group (tDMRG) \cite{tdmrg} this lack of efficiency 
has been traced back \cite{eff-dmrg} to a too
fast increasing of the entanglement entropy between parts of the whole system 
and the impossibility for a classical computer to store and manipulate such
large amount of quantum information. 

This observation partially moved the interest from the study of local
observables to the understanding of the evolution of the entanglement
entropy and in particular to its growth with time \cite{tee,cc-05}. 
Based on early results from conformal field theory \cite{cc-05,cc-07b}
and on exact/numerical ones for simple solvable 
model \cite{cc-05,exs} it is
widely accepted \cite{eff-dmrg} that the entanglement entropy 
grows linearly with time 
for a so called global quench (i.e. when the initial state differs globally
from the ground state and the excess of energy is extensive), while at most
logarithmically for a local one (i.e. when the the initial state has only a 
local difference with the ground state and so a little excess of energy). 
As a consequence a local quench is simulable by means of tDMRG, while a global
one is not.

However, despite this fundamental interest and a large effort of the
community, still analytic results are lacking. 
In this letter we fill this gap 
providing the analytic expression of the entanglement entropy at any time
in the limit of a large block for the XY chain in a transverse magnetic field
%The model is 
described by the Hamiltonian 
\be
H(h,\gamma)=-\sum_{j=1}^N\left[ \frac{1+\gamma}{4}\s^x_j\s^x_{j+1}+ 
\frac{1-\gamma}{4}\s^y_j\s^y_{j+1}+ \frac{h}2 \s^z_j\right],
\ee
where $\s_j^\alpha$ are the Pauli matrices at the site $j$.
Periodic boundary conditions are always imposed.
In spite of its simplicity, the model shows a rich phase diagram being critical
for $h=1$ and any $\gamma$ and for $\gamma=0$ and $|h|\leq1$, with the two
critical lines belonging to different universality classes.
The {\it block} entanglement entropy is defined as the Von Neumann entropy 
$S_\ell=-\Tr \rho_\ell\log \rho_\ell$, where $\rho_\ell=\Tr_{n\geq\ell}\rho$
is the reduced density matrix of the block formed by 
$\ell$ contiguous spins. In the following we will consider the quench 
with parameters suddenly changed at time $t=0$ from $h_0,\gamma_0$
to  $h,\gamma$. 
%(in the following the index $0$ will always refers to
%quantities related to the initial state).

Our main result is that, in the thermodynamic limit $N\to\infty$ and 
subsequently in the limit of a large block $\ell\gg1$, the time dependence of
$S_\ell(t)$ can be written as an integral over the momentum variable $\p$ 
\be
S_\ell(t)=t\!\! \int\limits_{2|\e'|t<\ell}\frac{d\p}{2\pi} 2|\e'| H(\cos\Delta_\p) +\
\ell\!\! \int\limits_{2|\e'|t>\ell}\frac{d\p}{2\pi} H(\cos\Delta_\p)\,,
\label{St}
\ee
where $\e'=d\e/d\p$ is the derivative of the dispersion relation 
$\e^2=(h-\cos\p)^2+\gamma^2\sin^2\p$ and represents the momentum dependent
sound velocity (that because of locality has a maximum we indicate as
$v_M\equiv \max_\p |\e'|$),
$\cos \Delta_\p=(h h_0-\cos\p (h+h_0)+\cos^2\p+\gamma\gamma_0\sin^2\p)/\e\e_0$ 
contains all the quench information \cite{foot1} and 
$H(x)=-[(1+x)/2\log(1+x)/2+(1-x)/2\log(1-x)/2]$. 

We first prove (\ref{St}) and then discuss its
interpretation and physical consequences.  The readers not interested in
the derivation can jump directly to the latter part.

{\it The method.} The entanglement entropy can be written in terms of a
block Toeplitz matrix \cite{Vidal,cc-05}. One first
introduce Majorana operators $
\Ma_{2l-1} \equiv \left( \prod_{m<l} \sigma_m^z \right) \sigma_l^x$ and
$\Ma_{2l} \equiv \left( \prod_{m<l} \sigma_m^z \right) \sigma_l^y$
and the correlation matrix $\Gamma^A_\ell$ through the relation
$\ev{\Ma_m\Ma_n}=\delta_{mn}+i{\Gamma^A_\ell}_{mn}$ with $1\leq m,n\leq \ell$, 
that is a block Toeplitz matrix
$$
\Gamma_\ell = \left[
 \begin{array}{ccccc}
\Pi_0  & \Pi_{1}   &   \cdots & \Pi_{\ell-1}  \\
\Pi_{-1} & \Pi_0   & &\vdots\\

\vdots&  & \ddots&\vdots  \\
\Pi_{1-\ell}& \cdots  & \cdots  & \Pi_0 
\end{array}
\right], ~~~ \Pi_l = \left[\begin{array}{cc}
-f_l    & g_l \\
-g_{-l} & f_l
\end{array}
\right]\,,
%\label{eq:GammaAL}
$$
with (as a straightforward generalization of \cite{cc-05})
\bea
g_l &=&  \int_{-\pi}^{\pi} \frac{d\p}{2\pi} e^{-i\p l} e^{-i\theta_\p}
(\cos \Delta_\p+i\sin \Delta_\p \cos 2\e_\p t)
\,,\nonumber\\
f_l&=& i\int_{-\pi}^{\pi} \frac{d\p}{2\pi} e^{-i\p l} \sin \Delta_\p \sin 
2 \e_\p t\,,
\label{eq:g2}
\eea
$\cos\theta_\p=-(h-\cos\p)/\e$, and 
$\sin\Delta_\p=-\sin\p[\gamma h_0-\gamma_0 h-\cos\p(\gamma-\gamma_0) ]/\e\e_0$.
The entanglement entropy is given by
\be
S_\ell=-\Tr \left[\frac{1+i\Gamma_\ell}2\log \frac{1+i\Gamma_\ell}2
\right]\,.
\label{S}
\ee
This trace can be numerically evaluated for finite $\ell$
as done in Ref. \cite{cc-05} in the Ising case ($\gamma=\gamma_0=1$) 
for $h,h_0\geq1$. A strong numerical evidence supports the fact that 
for $2 v_M t<\ell$, $S_\ell(t)$ increases linearly with time for large enough 
$\ell$, but this remained without any proof until now. 
It has also been argued that the limit $t\to\infty$ exists and $S_\ell$ 
saturates to a value proportional to $\ell$ (oppositely to the ground-state 
expectation where there is at most a logarithmic $\ell$ 
dependence \cite{cc-04,ent-rev,Vidal}) that has been calculated with a 
generalization of the Szego lemma \cite{cc-05}.
Eq. (\ref{St}) not only provides the proof for a strictly linear increasing
of $S_\ell(t)$ for $t< \ell/2v_M$, 
but gives also the complete time dependence.

{\it Proof of Eq. (\ref{St})}. 
Let us first sketch the strategy to prove Eq. (\ref{St}) and give only after
the technical details. 
The matrix $i \Gamma_\ell$ has the same eigenvalues of the $\ell\times \ell$ 
hermitian Hankel+Toeplitz matrices  $W_\pm=H\pm i T$, with 
$H_{k j}=g_{\ell+1-k-j}$ and $T_{k j}=f_{k-j}$.  
In fact, if $\vec{w}$ is an eigenvector of $W_\pm$ with eigenvalue $\omega$, 
then the vector $\vec{u}$ with elements $u_{2j-1}=w_j$ and 
$u_{2j}=\pm w_{\ell+1-j}$ is an eigenvector of $i\Gamma_\ell$ with 
eigenvalue $\mp \omega$.
We will show that $\Tr \,W^{2n}$ with $n$ integer satisfies a relation similar 
to Eq. (\ref{St}) with $H(x)$ replaced by the appropriate power.
Since Eq. (\ref{S}) can be written as an expansion in  $\Tr\, W^{n}$ with only
even powers, this implies Eq. (\ref{St}). 
Another ingredient is that in  Eq. (\ref{eq:g2}) for $\ell\gg1$ the term 
$e^{-i\theta_\p}$ is stationary being $\ell$ independent.
At this point a multidimensional integral for $\Tr\, W^{2n}$ (see below)
can be calculated with the stationary phase approximation, 
that is exact for $\ell\gg1$.

In order to get $\Tr{[W^n]}/\ell,\ n\in \mathbb{N}$ as $\ell\to \infty$ with 
a multidimensional stationary phase approximation (see e.g. \cite{wong}), 
we consider the Toeplitz and the Hankel symbols 
%$f(\p)$ and $g(\p)$ be the 
\bea
T_{k j}&=&\int_{-\pi}^{\pi}\frac{\mathrm{d}\p}{2\pi} e^{-i(k-j)\p}t(\p)\,,
\nonumber\\
H_{k j}&=&\int_{-\pi}^{\pi}\frac{\mathrm{d}\p}{2\pi} e^{-i(k+j-\ell-1)\p}h(\p)\, .
\eea
Each multiplication between two $H+T$ matrices involves sums like
\begin{multline}
\sum_{j=1}^\ell \exp\left[{i(j-(\ell+1)/2)(\p_1\mp\p_2)}\right]=\\
=\ell \int_{-1}^1{\mathrm{d}\xi_{12} \frac{\p_1\mp\p_2}{4\sin{\left(\frac{\p_1\mp\p_2}{2}\right)}}\cos{\left(\ell\xi_{12}\frac{\p_1\mp\p_2}{2}\right)}}\, ,
\end{multline}
where the sign is plus when an Hankel matrix is on the left and minus
otherwise. We can fix the sign to be minus, multiplying any 
Hankel symbol to the right by the parity operator acting as $Pf(x)P=f(-x)$. 
%Some difficulties arise 
When we close the chain with the trace operator, we cannot change the sign of 
the first symbol and if there is an odd number of $\mathrm{P}$,
%When we cannot fix the sign to be minus in the whole product (i.e. there is an 
%odd number of $\mathrm{P}$), 
the term does not contribute to the leading order, because the phase in the 
integrand is not stationary anymore. 
The $\xi$ dependence on the phase implies that all $\p_j$ variables are equal 
along the stationary curve. 
Note that all the significant terms have the same number of
Hankel symbols with reversed signs and with not. 
In fact, considering a tensor product of $n$ symbols inside a symmetric 
integration, it follows
(with $\simeq$ we always mean equal in the limit of large $\ell$) 
\be
(t+h \mathrm{P})^{\otimes n}\simeq\sum_k a_k^{(n,t)}h^{\otimes k}\otimes h_-^{\otimes k}+b_k^{(n,t)}h^{\otimes (k+1)}\otimes h_-^{\otimes k}\mathrm{P}\, ,
\ee
with \(h_-=\mathrm{P}h\mathrm{P}\), as can be straightforwardly proved by
induction.  
%Let us  prove this statement by induction on \(n\): for \(n=1\) it is a trivial assert, assuming the formula holds for \(\tilde{n}\) it is easy to check that
%\bea
%a_k^{(\tilde{n}+1,t)}&\sim& t\otimes a_k^{(\tilde{n},t)}+b_{k-1}^{\tilde{n},t}\,,\nonumber\\
%b_k^{(\tilde{n}+1,t)}&\sim& t\otimes b_k^{(\tilde{n},t)}+a_{k}^{\tilde{n},t}\, ,
%\eea
%and the proof is complete. 
%From what has already been proved and 
Using the parity $h^\ast(\p)=h(-\p)$, we have
that the Hankel symbol phase $e^{-i\theta_\p}$, 
which is not proportional to $\ell$ can be dropped.
%: the final result is not affected if we arbitrarily change
%integration variables in the slow oscillating part of the symbols. 
The Toeplitz symbol is odd $t(\p)=-t(-\p)$ so the symmetrized product
${(t(\p_1)+h(\p_1)\mathrm{P})}\otimes(t(\p_2)+h(\p_2)\mathrm{P})\simeq
t(\p_1)t(\p_2)+h(\p_1)h(-\p_2)$ leaves no parity terms, and if $n$ is odd
the whole integral asymptotically vanishes. 
Thus, from now on, we use the redefined symbols 
\bea
h(\p)&=&\cos\Delta_\p-i \sin\Delta_\p\cos 2\e t\,,\nonumber\\
t(\p)&=&\sin\Delta_\p \sin 2\e t\,,
\eea
which depend on the initial parameter only through %the angle 
$\Delta_\p$.
%, and the remaining $\theta_\p$ dependence is only through the energy $\e_\p$.
%Because we are interested in time $t\sim \ell$, we cannot omit to specify the
%variable dependence from the energies as we can do, on the other hand,
%considering terms like $\cos\Delta_\p$. 
Repeated application of the multiplication rule leads to
\be
\frac{1}{\ell}\Tr{W^{2n}}\simeq \frac{\ell^{2n-1}}{(4\pi)^{2n}}\int\limits_{C_{[-\pi,\pi]}^{(2n)}}\!\!\!\!\mathrm{d}^{2n}\p\!\!\!\!\!\!\int\limits_{C_{[-1,1]}^{(2n)}}\!\!\!\!\mathrm{d}^{2n}\xi
A\times B\,,
\ee
with
\bea
A&=&\prod_{j=1}^{2n}\cos\left[\ell\xi_{j}\frac{\p_j-\p_{j-1}}{2}\right]\,,\\
B&=&\prod_{j=1}^n \left[\cos^2\Delta_\p+\sin^2\Delta_\p\cos[2\e{(\p_{2j-1})}t
-2\e{(\p_{2j})}t]\right],\nonumber
\eea
%\begin{widetext}
%\be
%\frac{1}{\ell}\Tr{W^{2n}}\simeq\frac{\ell^{2n-1}}{(4\pi)^{2n}}\int\limits_{C_{[0,2\pi]}^{(2n)}}\!\!\!\!\mathrm{d}^{2n}\p\!\!\!\!\!\!\int\limits_{C_{[-1,1]}^{(2n)}}\!\!\!\!\mathrm{d}^{2n}\xi\ \underbrace{\prod_{j=1}^{2n}\cos\left[\ell\xi_{j}\frac{\p_j-\p_{j-1}}{2}\right]}_A\underbrace{\prod_{j=1}^n{\left(\cos^2\Delta+\sin^2\Delta\cos(2\e{(\p_{2j-1})}t-2\e{(\p_{2j})}t)\right)}}_B\, ,
%\ee
%\end{widetext}
where $C$ is the hypercubic domain. The product  $A$ can be moved inside the 
cos function turning it in a sum, because of the symmetry of the $\xi$ domain 
of integration with respect to $0$. 
There is a trivial integration along a direction in the $\xi$ domain since 
the integrand depends only on the difference between the $\xi$ variables, thus
$$
\int\limits_{C_{[-1,1]}^{(2n)}}\!\!\!\!\!\!\!\mathrm{d}^{2n}\xi\ A\simeq\int\mathrm{d}^{2n-1}\zeta\ \Omega(\zeta)\cos\left[\ell\sum_{j=1}^{2n-1}\zeta_{j}\frac{\p_j-\p_{2n}}{2}\right]\,,
$$
with 
$$
\Omega(\zeta)=\max[0,\min_j(1,1-\sum_{i=1}^j\zeta_i)+\min_j(1,1+\sum_{i=1}^j\zeta_i)]\, .
$$
The permutation symmetry of \(\p\) variables allows  to order the cos
products in $B$, so that we can introduce a set of spin variable
$\sigma_j\in\{-1,1\}$ and bring $B$ to the form 
\begin{multline}
B\simeq\sum_{k=0}^n \binom{n}{k}\frac{(\cos\Delta)^{2n-2k}(\sin\Delta)^{2k}}{2^k}\times\\
\times\!\!\!\sum_{\{\sigma_i\}_{i=1\dots k}}\!\!\!\!\cos{\left[\sum_{i=1}^k (2\sigma_i \varepsilon_{2i-1} t-2\sigma_i\varepsilon_{2i}t)\right]}\, .
\end{multline}
Interchanging the limit \(\ell\rightarrow \infty\) with the integration with
respect to the variable \(\p\equiv\p_{2n}\), the remaining
\((2n-2)\)-dimensional integral is easily solved by stationary phase methods. 
For each configuration of \(\{\sigma\}\) the Hessian determinant is $2^{2-4n}$ 
and the Hessian signature vanishes (%note that this is {\it} the fundamental
%property that allowed us to get the final result, and it 
this is the reason why in Eq. (\ref{St}) oscillations,
usually present in stationary phase calculations, are not present). 
On stationary points we finally have $\Omega=2$ on
time-independent terms and $\Omega=2(1-2|\e^\prime|t/\ell)$ otherwise. 
Of course the limit exists and the direct computation gives
\begin{multline}
\lim_{\ell\to \infty} \frac{\Tr{W^{2n}}}{\ell}=
\int_{-\pi}^{\pi}{\frac{\mathrm{d}\p}{2\pi}(\cos\Delta_\p)^{2n}}+\\
+\!\!\!\int\limits_{2|\e^\prime|t<\ell}\frac{\mathrm{d}\p}{2\pi}
\left(1-(\cos\Delta_\p)^{2n}\right)\left(1-2|\e^\prime|\frac{t}{\ell}\right)\,,
\label{pow}
\end{multline}
that is a ``Taylor'' expansion of Eq. (\ref{St}).
From Eq. \eqref{pow} we %not only have the time
%dependence of the von Neumann entropy, but 
also have the time dependence of all R\'{e}nyi entropies
$S_R={\log\Tr[\rho_l^\alpha]/(1-\alpha)}$: it is enough to replace
$H(\cos\Delta)$ with
${\log(|\cos^{2}\frac{\Delta}{2}|^\alpha+|\sin^{2}\frac{\Delta}{2}|^\alpha)/(1-\alpha)}$
in Eq. (\ref{St}).
On passing, we mention that our exact result obviously satisfies the bound for
$S_\ell(t)$ given in Ref. \cite{s2}.

{\it Description of the result}.
In Ref. \cite {cc-05} an interpretation of the time dependence of 
$S_\ell$ has been provided in terms of causality (later generalized to the 
correlation functions in \cite{qq}). 
The idea is simple: the initial state has a very high 
energy relative to the ground state of the Hamiltonian which governs the 
time evolution, and therefore acts as a source of quasiparticle excitations.
Particles emitted from different points (further apart than the
correlation length in the initial state) are incoherent, but pairs of
particles moving to the left or right from a given point are entangled. 
Thus $S_\ell(t)$ should just be proportional to the number of 
coherent particles that emitted from any point reach one a point in $[0,\ell]$
and the other the remainder of the system. Since there is a maximum speed 
for these excitations $v_M$, this implies the linear growth for $2v_Mt<\ell$
and saturation for very large times. 

However, only in the conformal case when $\e'$ does not depend on the momentum
because of the linear dispersion relation, this scenario makes
quantitative predictions on the time evolution, else the rate of production of
particles $f(p',p'')$ is an unknown function of
the Hamiltonian parameters both before and after the quench. 
The comparison of Eq. (\ref{St}) with the general one (Eq. (4.2) in
\cite{cc-05}) allows to identify 
$f(p',p'')$ with $\delta(p'-p'') H(\cos\Delta_{p'})$. 
\begin{figure}[t]
\includegraphics[width=0.45\textwidth]{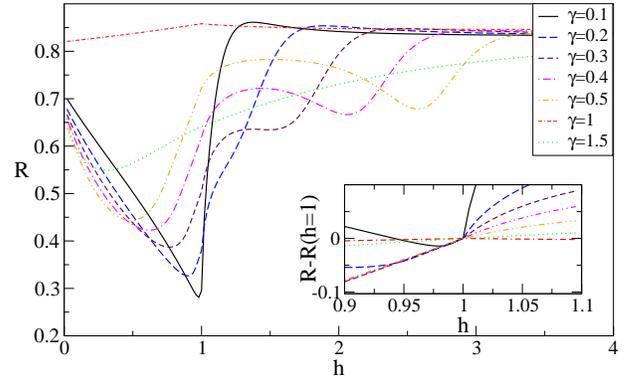}
\caption{\label{figR}
R defined in Eq. (\ref{RR}) for quenches from $(h_0=10, \gamma_0=2)$. 
The inset shows the singular behavior in the neighborhood of $h=1$.}
\end{figure}
We can also easily read from our result the value of the ratio
\be
R\equiv\frac{(\partial S_A/\partial t)_{t<t^*}}{2v_M
(\partial S_A/\partial\ell)_{t\gg t^*}}=
\frac{\int_{-\pi}^{\pi}d\p |\e'| H(\cos\Delta_\p)}{
v_M \int_{-\pi}^{\pi}d\p H(\cos\Delta_\p)}\,,
\label{RR}
\ee
that results to be the average of the absolute value of speed of the 
sound on the $H(\cos\Delta_\p)$ distribution. 
$R$ as function of the quench parameters is shown in Fig. \ref{figR}. 
It is not analytic at the quantum critical point $h=1$, as a trivial
consequence of the non-analyticity of its building blocks (i.e. $\e$,
$\Delta_\p$). 
However it is clear from the inset that such non-analyticity is so weak 
that is unrealistic to say that the out-of-equilibrium behavior of
entanglement entropy is sensitive to the phase transition.

From Eq. (\ref{St}) we also have the large time corrections to the 
asymptotic result. Since $H(\pm1)=0$ with a log singularity, when the 
zero-velocity mode giving the large $t$ behavior is at $\p=\pm\pi$ (as 
e.g. for $h>1$), one has that
the first correction is $\propto\ell^4\log t/t^3$, whereas when there are 
zero-velocities not at the border of the Brillouin zone, where $H(x)$ is 
finite, the leading correction is $\propto\ell^2/t$.

For $t=\infty$ only the second term in Eq. (\ref{St}) contributes
to the entropy that thus is extensive. As already noticed in
Ref. \cite{cc-05} this is the same result at finite {\it large} temperature 
$\beta_{\rm eff}$.  
An interesting question is whether this effective temperature is observable
independent as found in the conformal case \cite{qq} or instead depends on 
the operator and so would not be a well-defined concept. 
Further checks of this point are mandatory to avoid speculations.

As a consequence of $H(x)\leq H(0)=\ln 2$, we have $S_\ell(t)\leq \ell \ln 2$
for any time, a bound that is just the maximum entanglement allowed
by the dimension of the Hilbert space.
In Ref. \cite{cc-05} it was noticed that the various curves for the quench
from $h_0=\infty$ to any $h$ apparently collapse on a single curve when 
rescaling $S_{\ell}(t)$ to $S_{\ell}(\infty)$. 
From Eq. (\ref{St}), %it is easy to check that 
this is exactly true for $|h|\leq1$, but only approximately otherwise. 
%A final observation is that 
Finally, the $t=\infty$ result is symmetric under the
exchange $(h,\gamma)\leftrightarrow (h_0,\gamma_0)$,
because the asymptotic result only depends on
$\cos\Delta_\p$ that does not distinguish between initial and final values.

\begin{figure}[t]
\includegraphics[width=0.48\textwidth]{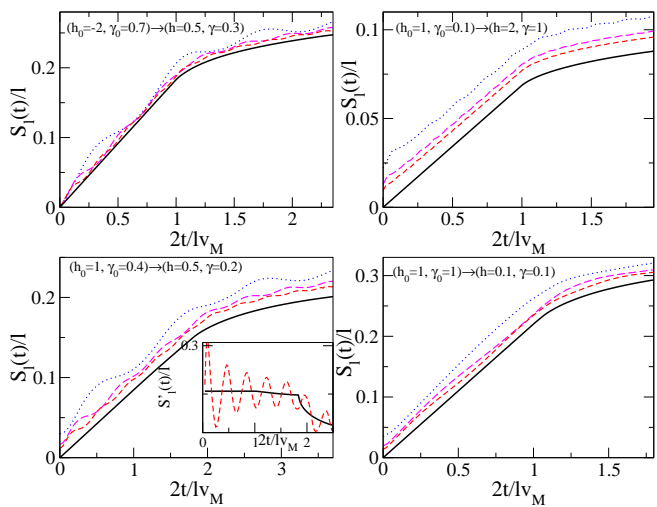}
\caption{\label{figSt}Time evolution of the entanglement entropy 
  $S_\ell(t)/\ell$ for several quenches and $\ell$. 
  The straight line is the leading asymptotic result for  large $\ell$. 
  The inset in the bottom-left graph shows the derivative with  respect to
  time of $S_\ell(t)$ for $\ell\to\infty$ and the numerical derivative 
  for $\ell=90$.}
\end{figure}

{\it Finite $\ell$}.
The matrix representation (\ref{S}) allows for the numerical calculation of
$S_\ell$ for finite and relatively large $\ell$.
Some results are reported in Fig. \ref{figSt}, where we choose those quench
parameters that make the finite $\ell$ effects more relevant.
Increasing $\ell$ the results always approach Eq. (\ref{St}),
showing unambiguously its correctness, 
but there are peculiar and interesting finite $\ell$ effects.
The most evident effect is the oscillation of $S_\ell(t)$. 
These oscillations have been generically seen in numerical studies also on more
complicated models \cite{exs}, but they are absent in 
the Ising model for $h,h_0\geq1$ \cite{cc-05}: oscillations can only be
present when there is a second {\it local} maximum of $|\e'|$. 
The data (also for cases not shown in the figure) provide a strong evidence
that the first non-oscillating correction at order $O(\ell^0)$ is positive and
time independent.

In the bottom-left plot in Fig. \ref{figSt} 
the most unexpected effect is shown. 
For the quench $(h_0=1,\gamma_0=0.4)\to (h=0.5,\gamma=0.2)$,
it seems that the linear regime of $S_\ell(t)$ continues
after $t^*=\ell/2v_M$. However, looking at the derivative (inset) one 
realizes that it is not exactly constant, since it slightly bends 
at $t^*$. This happens because for this peculiar quench 
the maximum velocity mode carries very little information, and so a stronger
non-analyticity is present at a local maximum of the velocity 
smaller than $v_M$. This effect is pronounced every time 
that $h_0 \gamma\sim h \gamma_0$, with $|h|, |\gamma|,|h_0|,|\gamma_0|<1$,
because of the functional form of $\Delta_\p$. 
This anomalous behavior is important because it is nowadays common to extract
the speed of propagation of information from $t^*$.
Every time this effect is present, this procedure gives the wrong
answer. For example, we plotted in the inset of Fig. \ref{figSt} the numerical
derivative of $S_\ell(t)$ for $\ell=90$ (a value hardly reached in
non-equilibrium simulation). It is evident that at $t^*$ there is no trace of
the non-analyticity. Relying on these results one would have obtained a
value of $v_M$ that is almost half of the real one.

{\it Conclusions}. The non equilibrium time-evolution of $S_\ell(t)$ for the
XY chain seems to encode most of the features that have been observed
numerically in other contexts \cite{exs}. It is then natural to wonder whether
slight modifications of Eq. (\ref{St}) can be true in more complicated 
situations 
and not only for models mapped to free fermionic theories as the present one.

{\it Acknowledgments.}
PC benefited of a travel grant from ESF (INSTANS program).

\end{document}